\documentclass[fleqn,usenatbib,useAMS]{mnras}
\usepackage{graphics,epsf}
\usepackage{amsmath}                % American Mathematical Society package
\usepackage{amsfonts}               % American Mathematical Society fonts
\usepackage{amssymb}                % American Mathematical Society symbol
\usepackage{epsfig}                 % EPS figures
\usepackage{graphicx}

\usepackage{xcolor}
\definecolor{redak}{rgb}{0.9,0.15,0.05}
\def \rmModot{\,\rm{M_\odot}}
\def \rmRodot{\,\rm{R_\odot}}
\def \rmLodot{\,\rm{L_\odot}}

\title[Novae Induced by Wind Accretion]{Simulations of Multiple Nova Eruptions Induced by Wind Accretion in Symbiotic Systems}

\author[Y. Hillman and A. Kashi]{
Yael Hillman$^{1}$\thanks{Yael Hillman e-mail: \href{mailto:yaelhi@ariel.ac.il}{yaelhi@ariel.ac.il}}
and
Amit Kashi$^{1}$\thanks{Amit Kashi e-mail: \href{mailto:kashi@ariel.ac.il}{kashi@ariel.ac.il}}
\\
$^{1}$Department of Physics, Ariel University, Ariel, POB 3, 4070000, Israel \\
}

\date{Accepted 2020 November 12. Received 2020 November 11; in original form 2020 September 06}

\pubyear{2020}

\begin{document}
\label{firstpage}
\pagerange{\pageref{firstpage}--\pageref{lastpage}}
\maketitle

\begin{abstract}
We use a combined binary evolution code including dynamical effects to study nova eruptions in a symbiotic system.
Following the evolution, over $\sim10^5$ years, of multiple consecutive nova eruptions on the surface of a $1.25\rmModot$ white dwarf (WD) accretor,  
we present a comparison between simulations of two types of systems.
The first is the common, well known, cataclysmic variable (CV)
system in which a main sequence donor star transfers mass to its WD companion via Roche-lobe overflow. The second is a detached, widely separated, symbiotic system in which an asymptotic giant branch donor star transfers mass to its WD companion via strong winds. For the latter we use the Bondi-Hoyle-Lyttleton prescription along with orbital dynamics to calculate the accretion rate. We use the combined stellar evolution code to follow the nova eruptions of both simulations including changes in mass, accretion rate and orbital features.
We find that while the average accretion rate for the CV remains fairly constant, the symbiotic system experiences distinct epochs of high and low accretion rates. The examination of epochs for which the accretion rates of both simulations are similar, shows that the evolutionary behaviors are identical. We obtain that for a given WD mass, the rate that mass is accreted ultimately determines the development, and that the stellar class of the donor is of no significance to the development of novae.
We discuss several observed systems and find that our results are consistent with estimated parameters of novae in widely separated symbiotic systems.
\end{abstract}


\section{Introduction}\label{sec:intro}
Novae are known to occur in cataclysmic variable (CV) systems \cite[]{Warner1995}, comprising a white dwarf (WD) and a less evolved, low mass main sequence (MS) companion \cite[]{Kraft1964}, where the typical orbital period ($P\rm_{orb}$) is of the order of several hours. Hydrogen rich matter is transferred from the less evolved star (the donor) via Roche-lobe (RL) overflow and piled up on the surface of the WD. 
Eventually, a critical amount of mass will accumulate, raising the pressure and temperature at the base of the accreted envelope enough to ignite the hydrogen in a thermonuclear runaway (TNR) process \cite[]{Starrfield1972}, violently ejecting the accreted envelope. This is observed as a sudden brightening \cite[]{Pay1957}, of order $\sim10^4~\rmLodot$, followed by a slow decline \cite[e.g.,][]{Prialnik1986}. 

In wide symbiotic systems comprising a WD and an evolved swollen giant, typical orbital periods are of the order of years. If the separation is sufficiently wide, the stellar components will be too far apart to allow RL overflow (RLOF) \cite[]{Murset1999,Skopal2019}, however, part of the mass that is blown away from the giant via strong stellar winds may be captured by the WD's gravity. 
\cite{Skopal2019} explains that sudden brightenings in symbiotic systems are not nearly as bright as nova eruptions in CVs and may occur via two possible processes; due to accretion alone, which may cause a brightening of $1$--$100\rmLodot$ originating from instabilities in the accretion disk (e.g., EG And and 4 Dra \cite[]{Skopal2005}); or non-eruptive systems, such as, RW Hya and SY Mus that brighten due to more or less stable nuclear burning of the accreted hydrogen \cite[e.g.,][]{Paczynski1980,Mikolajewska2008} on the surface of a low mass WD as a result of a certain rate of accretion \cite[e.g.,][]{Shen2007,Wolf2013}, which may lead to a luminosity of $\sim10^3\rmLodot$ \cite[e.g.,][]{Muerset1991}. However, the possibility of stable nuclear burning in such systems as a result of hydrogen accretion onto the surface a WD has been questioned in recent years, due to the requirement of a constant and accurate rate of accretion  \cite[e.g.,][]{Starrfield2012a,Idan2013,Hillman2015}. 
Some symbiotic systems show weak eruptions of 1--3 $M_v$ that last for a few months to a few years where the WD remains at constant luminosity. \cite{Mikolajewska2008} explains this as activity in the accretion disk. 

Symbiotic novae are novae that occur in symbiotic systems. For example, AG Peg and RT Ser (which have a normal M giant donor --- S-type symbiotic systems) or V1016 Cyg and HM Sge (which have a Mira variable donor --- D-type symbiotic systems) \cite[]{Mikolajewska2008}. The manner of mass transfer from the donor to the WD may be via RLOF or via wind, while the opinions differ as to which mechanism is preferred. For example, 
\cite{Rudy1999} suggests that one of the basic differences between novae in CVs and in symbiotic systems is in the respect that the transfer of mass arises from stellar wind rather than from RLOF, whereas \cite{Strope2010} claims that the orbital periods of most of these types of systems are relatively short and therefore their binary components are close enough for the mass to be transferred via RLOF as in CVs.

Evolved giants are characterized by strong stellar winds driven by radiation pressure on the outer layers of their atmospheres \cite[]{Baud1983,Vassiliadis1993,Dorfi1998}. Stars on the red giant branch (RGB) experience relatively steady winds (i.e., mass loss), whereas stars on the asymptotic giant branch (AGB) have a more complex wind behavior. AGB stars undergo fusion in alternating hydrogen and helium overlaying shells \cite[]{Iben1975,Sugimoto1978,Renzini1981,Iben1983}.  Burnt hydrogen falls onto the helium shell below and replenishes it on a timescale of $10^4-10^5$ years, until a critical amount of helium is accumulated, triggering a helium flash \cite[]{Dorfi1998,Piovan2003}. This causes the star to inflate, increases the wind rate (thus causing epochs of higher and lower wind rates) and cools the overlaying hydrogen shell, hence quenching the fusion. When the flash subsides, the envelope recedes, allowing the temperature in the outer hydrogen shell to rise again, thus beginning another epoch of hydrogen burning.

In wide symbiotic binaries for which wind from the giant is captured by the WD, although this transfer of mass does not depend on RLOF, the binary separation is still one of the key players in determining the rate of mass captured by the WD. One of the main attributes affecting the separation is  drag force to which the WD is subject while moving within the wind expelled from the donor. Another attribute is the fact that due to the strong winds typical of stars on the AGB \cite[]{Baud1983,Dorfi1998}, mass is lost from this type of system much faster than in CVs. Both of these attributes have a non-negligible affect on the separation change. In addition, the wind rate changes due to the inflation and contraction that are typical of an envelope of an AGB star (as described above), which has a direct effect on the rate of mass accretion onto the WD.

Here we explore a path that produces TNR triggered multiple nova eruptions in wide symbiotic systems for which the mass transfer from the donor to the surface of the WD is not due to RLOF but rather from the expelled wind from an AGB star. 
The expelled mass that is captured by the gravity of the companion WD eventually accumulates on its surface and produces a nova eruption via a thermonuclear runaway (TNR), as in CVs. Our goal is to characterize these nova eruptions and compare their behavior with novae in CVs.  The accretion rate ($\dot{M}\rm_{acc}$) in CVs is one of the key parameters that determine novae features (e.g., timescales, accreted and ejected masses etc.) \cite[]{Shara1986,Kovetz1988,Livio1988,Ritter2000,Toonen2014}. \cite{Hillman2020} have shown that $\dot{M}\rm_{acc}$ changes, not just during a cycle of accretion, but it evolves over $\sim10^9$ years until the donor is almost entirely eroded. They demonstrated that the separation ($a$) is the most important parameter that determines the accretion rate. 

We present a self-consistent long-term simulation of a WD producing multiple consecutive nova eruptions due to accretion from the strong wind of an AGB star.We follow the evolution and the timescales of parameters such as the accretion rate ($\dot{M}\rm_{acc}$), the stellar masses (the WD mass, $M\rm_{WD}$, and the AGB donor mass, $M\rm_{D}$), the separation ($a$) and the orbital period ($P\rm_{orb}$), as well as the long-term behavior, and we compare our findings with novae in CVs. Our computing methods are detailed in \S\ref{sec:methods}, followed by a description of our models. We present our results in \S\ref{sec:results}, discuss their implications in \S\ref{sec:discussion} and conclude in \S\ref{sec:conclusions}.

\section{Methods}\label{sec:methods}
\subsection{Calculations}\label{sec:methods-calcs}

The calculations were carried out by adapting the combined code for simulating the long-term evolution of a binary system producing consecutive nova eruptions via RLOF (described in detail in \cite{Hillman2020}). In this code the donor component is simulated by using a hydrostatic stellar evolution code \cite[]{Kovetz2009} designed to follow the evolution of a star from pre-MS all the way through to a WD. This data is used in a hydrodynamic Lagrangian nova evolution code designed to follow the evolution of hundreds of thousands of nova cycles of accretion and eruption on the surface of a WD \cite[]{Prikov1995,Epelstain2007,Hillman2015}. The combined code recalculates the accretion rate onto the WD at each timestep by accounting for orbital momentum change due to magnetic braking (MB) and gravitational radiation (GR) using \cite[]{Paxton2015}:

\begin{equation}\label{eq:JMB}
\dot{J}_{MB}=-1.06\times 10^{20}M_{D}R_{D}^4P\rm_{orb}^{-3}
\end{equation}  
\begin{equation}\label{eq:JGR}
\dot{J}_{GR}=-\frac{32}{5c^5}\left(\frac{2\pi G}{P\rm_{orb}}\right)^\frac{7}{3}\frac{(M_{D}M_{WD})^2}{(M_{D}+M_{WD})^\frac{2}{3}}
\end{equation}

and for mass lost from the system during each nova eruption by calculating the separation change using:
\begin{equation}\label{eq:delta_a}
\Delta{a}=2a\left(\frac{m{\rm_{ej}}-m{\rm_{acc}}}{M_{WD}}+\frac{m\rm_{acc}}{M_{D}}\right)
\end{equation}

where $\dot{J}_{MB}$ and $\dot{J}_{GR}$ are the change in orbital angular momentum due to magnetic braking and  gravitational radiation respectively, $M_{WD}$ and $M_{D}$ are the masses of the WD and the donor respectively, $R_{D}$  is the radius of the donor, $P\rm_{orb}$ is the orbital period, and $m\rm_{acc}$ and $m\rm_{ej}$ are the accreted and ejected masses of the previous cycle respectively.
 
In order to accommodate the code for the accretion of wind from an AGB star we made the following two major adjustments. The first, involves determining the accretion rate onto the surface of the WD by using the Bondi-Hoyle-Littleton (BHL) prescription \cite[]{Hoyle1939,Bondi1944} (rather than by RLOF) which determines the rate of mass accretion onto an object that is moving in a cloud of gas, e.g., a single star in the interstellar medium. The method uses the density of the gas and the relative velocity between the object and the gas, as well as the mass and gravitational pull of the object.

We adopt the prescription from \cite{Kashi2009} for obtaining the time dependent rate of mass accretion onto the surface of the WD, while setting the eccentricity to zero and the acceleration wind model parameter 
to unity. Their method employs the long standing BHL accretion prescription together with the CAK model \cite[]{Castor1975} for radiation driven wind of the donor star, and taking into account the relative motion between the stars.
We obtain the relative velocity ($v_w$):

\begin{equation}\label{eq:vw1}
v_{w1}=v_\infty+\frac{R_{D}}{a}(v_s-v_\infty)
\end{equation}
\begin{equation}\label{eq:vorb}
v{\rm_{orb}}^2=\frac{G(M_{WD}+M_{D})}{a}
\end{equation}
\begin{equation}\label{eq:vw}
v_w=(v{\rm_{orb}}^2+v_{w1}^2)^\frac{1}{2}
\end{equation}
where $v_{w1}$ is the wind speed at a distance $a$, $v_s$ is the sound velocity at the surface of the donor star, we take the terminal wind velocity, $v_\infty$, to be $20~\rm{km~s^{-1}}$ and $v\rm_{orb}$ is the orbital velocity of the WD. The accretion rate  calculated by using the BHL method is given as \cite[]{Bondi1944}:
\begin{equation}\label{eq:mdotacc}
{\dot{M}\rm_{acc}}=\frac{4\pi G^2M_{WD}^2\rho_w}{v_w^3}
\end{equation}

Where  $\rho_w$ is the density of the wind at the accretor, calculated from the wind rate ($\dot{M}_w$) using
\begin{equation}\label{eq:rho_w}
\rho_w=\frac{\dot{M}_w}{4\pi a^2 v_{w1}}.
\end{equation}
A detailed elaboration on this calculation method, and its adaption for obtaining the accretion rate from wind, is explained in \cite{Kashi2009}. We note that the selection of the wind acceleration parameter has only a small effect as the wind reaches the WD at close to its terminal velocity.

The second modification that we incorporated stems from the nature of the WD being embedded in a cloud of mass that has been blown away from the AGB star. Accretion from wind introduces the need for accounting for angular momentum loss due to drag. This is because, in contrast with accretion from an un-evolved (red dwarf) star, for which mass loss via stellar wind is negligible, AGB stars expel wind at rates that may reach an order of $\sim10^{-7}-10^{-5}\rmModot\rm{yr}^{-1}$ \cite[e.g.,][]{Kenyon1987,Chomiuk2012b}. This high rate is what actually enables using the BHL prescription described above --- by treating the WD as moving through a cloud of gas --- but this scenario inevitably introduces drag. Therefore, we have additionally included in our calculations the orbital momentum loss due to drag force on the WD. 
The drag force on a stellar object in a cloud of gas is given as \cite[]{Alexander1976}:

\begin{equation}\label{eq:Drag}
D=\pi r_{\rm{acc}}^2 \rho_w v_w^2
\end{equation}
where $r_{\rm{acc}}=2GM_{WD}/v_w^2$ is the accretion radius of the WD \cite[]{Bondi1944}.

Taking these effects into account allows us to follow dynamical changes throughout the system's evolution, and monitor how the orbital separation, accretion rate onto the WD and other parameters vary with time. Merging the dynamics code with the combined code of stellar and nova evolution allows a full study of the wind accretion problem, taking into account all the major effects that influence the evolution of a symbiotic system.

\subsection{Models}\label{sec:methods-models}

We compare between two models; a RLOF model and a wind accretion model. Both simulations begin with given initial stellar masses and separation as input parameters. The initial separation for the CV RLOF model is determined by setting $R_{D}{=}R_{RL}$, i.e., the donor is exactly filling its RL.
The initial separation for the wind model is chosen as far enough apart to ensure that throughout the evolution the donor does not fill its RL. The radius of the donor (AGB) star at the onset of the simulation is of order $\sim2{\times}10^2\rmRodot$ --- much smaller then the chosen separation (see Table \ref{tab:mdls}).

The initiation of the wind simulation is at the evolutionary point when the donor star first experiences strong winds (${\gtrsim}10^{-6}M_\odot \rm{yr}^{-1}$) which occurs in the AGB phase. 
The mass of the donor, which was chosen to be $1.0\rmModot$ at the onset of the simulation, requires the progenitor star to have had the zero age MS (ZAMS) mass of $1.35\rmModot$. For consistency, the donor of the RLOF model was chosen to be $1.0\rmModot$ as well, while it is a ZAMS star. The WD masses were chosen as $1.25\rmModot$ for both models and the composition of the accreted matter in both models is solar.

From these initial points the simulations run continuously until either the donor is eroded down to ${\lesssim}0.1\rmModot$ (for the CV RLOF model) or the donor becomes a WD (for the wind model).
These initial parameters are summed in Table \ref{tab:mdls}. 
\begin{table}
	\begin{center}
		\begin{tabular}{|c|c|c|c|c|}
			\hline
			{model}&{$M_{WD}[\rmModot]$}&{$M_{D}[\rmModot]$}& {$a[\rmRodot]$}& {$P\rm_{orb}$}\\
			\hline
			{WIND}&{1.25}& {1.0 AGB}&{3.40e3}&{42.0 [yr]}\\
			
			{RLOF}&{1.25}& {1.0 MS}&{2.55e0}&{7.54 [hr]}\\
			\hline
		\end{tabular}
		\caption{The models - initial. The wind model donor star is a giant on the AGB, its zero age MS mass is $M_{ZAMS}{=}1.35 \rmModot$. The RLOF donor star is a ZAMS star}\label{tab:mdls}
			\end{center}
\end{table}

\begin{table*}
	\begin{center}
		\begin{tabular}{|c|c|c|c|c|c|c|}
			\hline
			{model}&{$M_{WD,f}[\rmModot]$}& {$M_{D,f}[\rmModot]$}& {$a_f[\rmRodot]$}& {${P_{\rm{orb}},_f}$}&{cycles}&{$t{\rm_{total}}[\rm{yr}]$}\\
			\hline
			{Wind}&{1.2503}& {0.60 WD}&{3.22e3}&{42.6 [yr]}&{1924}&{2.6e5}\\
			
			{RLOF}&{1.2499}& {0.99 MS}&{2.55e0}&{7.53 [hr]}&{280}&{2.6e5$\rm^{a}$}\\
			\hline
		\end{tabular}   
		\caption{The models - final. $\rm^{a}$The evolution of the RLOF model is shown for a period of time matching the total evolution time of the wind model.}\label{tab:mdls1}
	\end{center}
\end{table*}

We note that in comparing between models, a useful technique would be to set as many parameters as possible as identical, ideally changing only one, and examining its effect on the development. Here we compare between two very different types of systems, and consequently there are a limited number of parameters that can be initially set as identical. Therefore we chose identical values for the initial binary masses for both types of systems, in an effort to minimize the free parameters of this comparison, even though the donor of the two types of systems are at different evolutionary phases.

\section{Results}\label{sec:results}
\subsection{Binary behavior}\label{sec:bin}
We find substantial timescale differences between the evolutionary paths of the two types of systems. While the donor of the RLOF model becomes eroded down to ${\lesssim}0.1\rmModot$ on the order of a few Gyr, the donor of the wind model becomes a WD in a fraction of this time --- of order $\sim10^5$ years --- thus ending the possibility for mass transfer and hence for novae. Therefore in order to compare between the two extremely different scenarios, we present results of the RLOF model for a period of time matching the total time of the wind model evolution. Final values corresponding to the input parameters are summed in Table \ref{tab:mdls1}. In addition we present a comparison over the number of nova eruptions. We find that regardless of which of these scales we examine, the models have entirely different paths, as we explain below.

\begin{figure*}%[!ht]%
	\begin{center}
		{\includegraphics[trim={0.0cm 0.2cm 0.0cm 1.3cm},clip ,
			width=1.0\columnwidth]{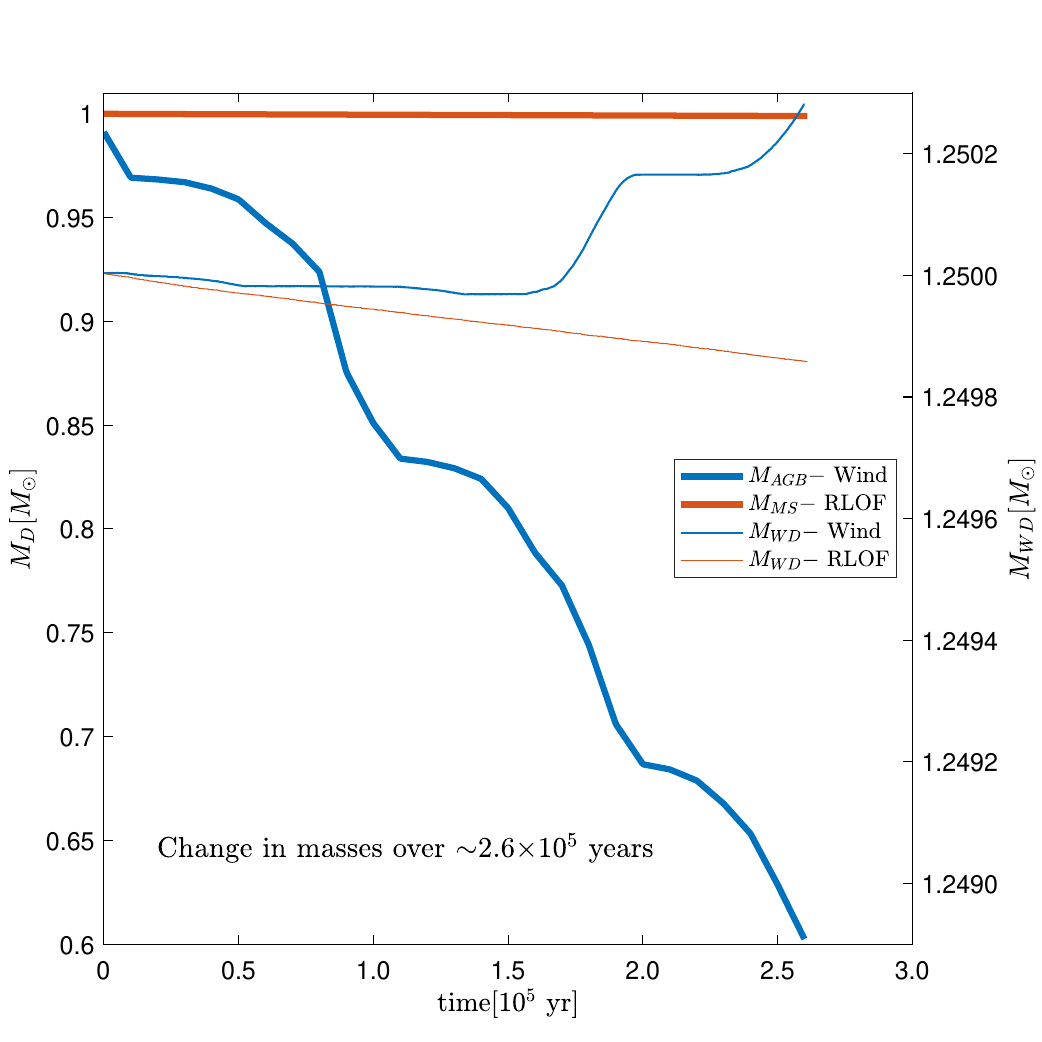}}
		{\includegraphics[trim={0.0cm 0.2cm 0.0cm 1.3cm},clip ,
			width=1.0\columnwidth]{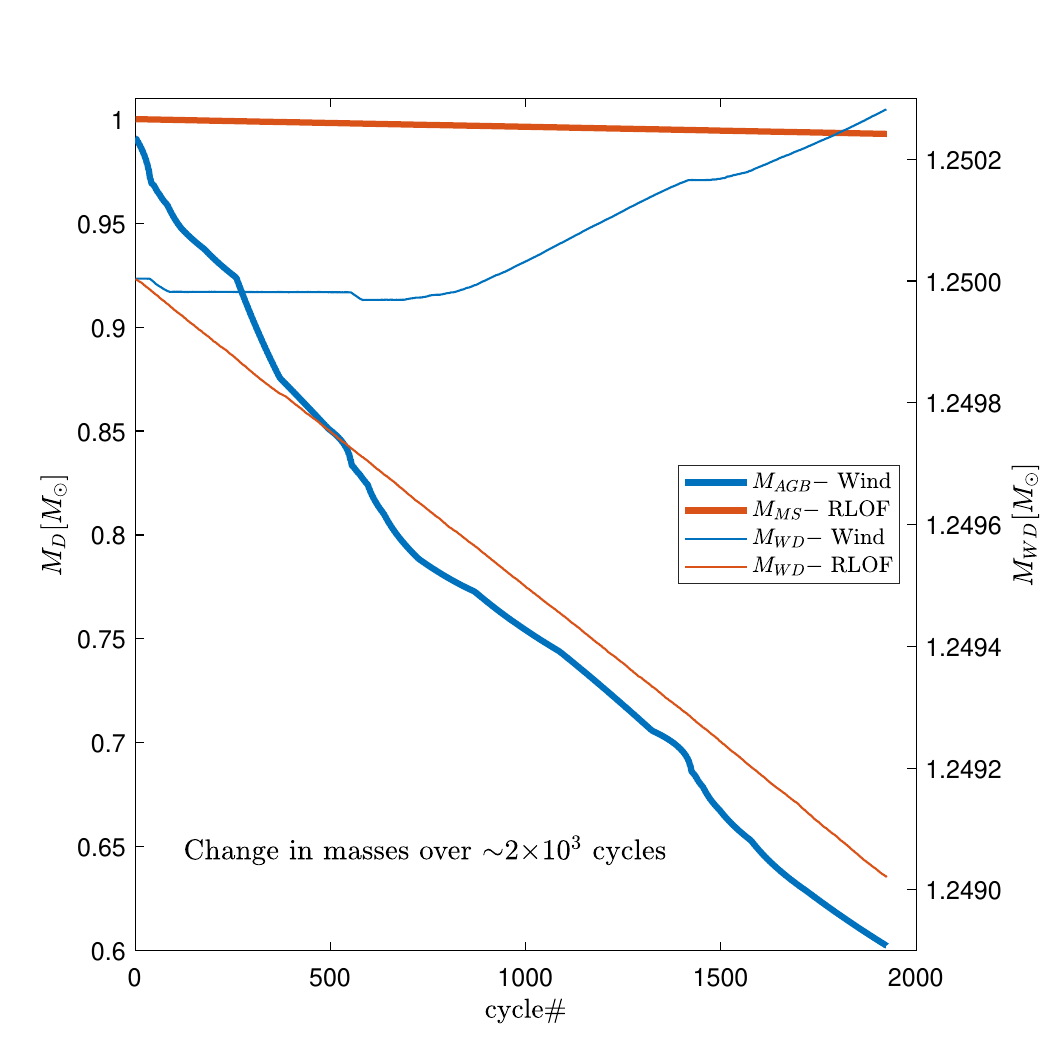}}
		\caption{The evolution of the stellar masses of both models on two perspectives: Both panels include the entire wind evolution ($2.6\times10^5$ years and $\sim2000$ cycles, blue). The left panel additionally shows the RLOF model evolution (red) over $2.6\times10^5$ years (which is 280 cycles) and the right panel additionally shows the RLOF model evolution  (red) over  $\sim2000$ cycles (which is $\sim1.8\times10^6$ years). 	 \label{fig:masses}}
	\end{center}
\end{figure*}

Figure \ref{fig:masses} shows the change in mass of the binary components vs. time (left panel) and vs. number of eruption (right panel). Although both models begin with the same masses, the systems' evolutionary paths are entirely different from each other. Over $\sim2.6\times10^5$ years the AGB star had lost a total of $\sim0.4\rmModot$ due to wind. Most of this wind was lost from the system, while a fraction of it was captured by the gravity of its companion WD, which underwent nearly 2000 periodic nova cycles of accretion and eruption. Most of the mass captured by the WD is periodically ejected during the nova eruptions.
The total net mass change of the WD during this time is $+\sim2.8\times10^{-4}\rmModot$ --- an average of $\sim1.4\times10^{-7}\rmModot$ mass \textit{growth} per eruption. A positive mass change means the average accretion rate must be relatively high, of order larger than a few times $\sim10^{-8}\rmModot \rm{yr}^{-1}$ \cite[]{Idan2013,Wolf2013,Hillman2015}. We note that these authors examined constant accretion rates, while the accretion rate of our wind model can by no means be considered constant, as will be discussed later.
 
In contrast with mass change behavior in the wind model evolution, over the same time period, both stellar masses of the RLOF model decrease monotonically, albeit not nearly by the same amounts of mass nor in the same fashion. All the mass that is lost from the donor is transferred to the WD which, as in the wind model, undergoes periodic nova cycles of accretion and eruption. Allowing the model to run indefinitely would result with the donor eroded down to $\lesssim0.1\rmModot$. However this evolutionary process is a few Gyrs long.
 When addressing only the first  $\sim2.6\times10^5$ years of evolution (to match the evolutionary time of the wind model) the total mass lost from the MS donor is merely $\sim1.06\times10^{-3}\rmModot$. This mass was accreted onto the WD at a slower average rate 
 than in the wind model, resulting in less nova cycles --- a total of only 280 cycles in this time period --- and a total net mass \textit{loss} of $\sim1.42\times10^{-4}\rmModot$ from the WD due to these nova eruptions.
 
Since using an identical time span yields a comparison between different amounts of cycles (280 RLOF model cycles vs. 1924 wind model cycles) we show as a test, an additional perspective in the right panel of Figure \ref{fig:masses}, where we compare the first 1924 eruptions of the RLOF model with the entire wind evolution, i.e., the same amount of cycles for both models. This perspective allows the examination of the evolution \textit{per eruption} rather than over a period of time. Of course, doing so, means that the time can not the same for both models --- the RLOF model requires nearly seven times longer ($\sim1.8\times10^6$ years) to complete this number of cycles, but, it is clear from the right panel of Figure \ref{fig:masses} that the trend remains the same --- the RLOF model's stellar components continue to lose mass monotonically, straying away from the wind model mass curves. This test removes any doubt that the evolutionary paths may be different due to less eruptions, since even when looking at the same amount of eruptions, the models do not develop in the same manner. The ultimate reason that the WD mass develops differently in the two models is the rate that the mass is accreted onto the WD, which we later explain.

\begin{figure}%[!ht]%
	\begin{center}
		{\includegraphics[trim={0.0cm 0.2cm 0.0cm 1.2cm},clip ,
			width=0.95\columnwidth]{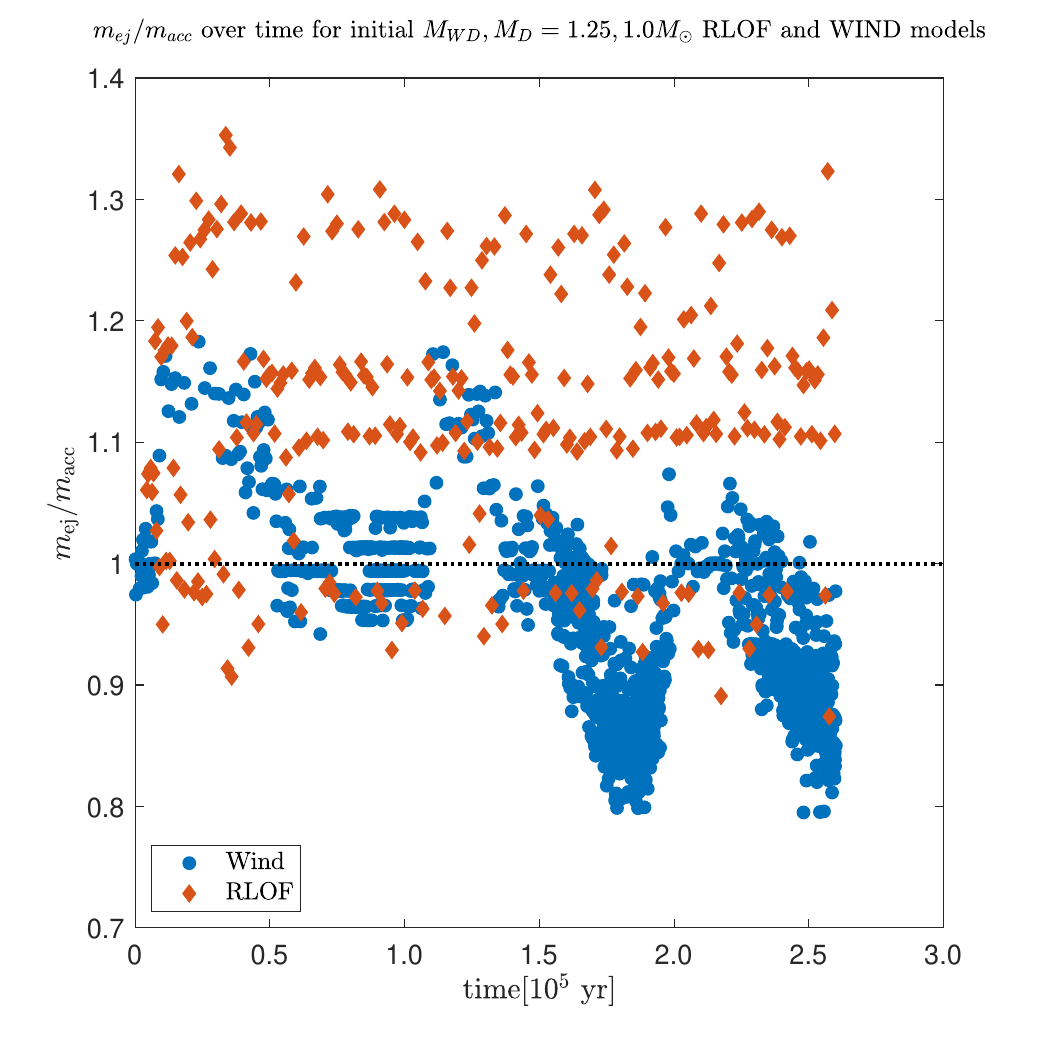}}
		\caption{The evolution of the ratio of the ejected mass to the accreted mass per cycle for the wind model (blue) and the RLOF model (red). \label{fig:mej_over_macc}}
	\end{center}
\end{figure}

To compliment Figure \ref{fig:masses} we show in Figure \ref{fig:mej_over_macc} the ratio ejected to accreted mass (${m\rm_{ej}}/{m\rm_{acc}}$) per nova cycle. Everything under the horizontal dotted line marking '1' contributes to mass growth, while everything above it leads to mass loss. The ratio for most of the RLOF nova eruptions are roughly evenly distributed in the range $\sim1.15\pm0.2$, meaning that most of the nova eruption end in a net WD mass reduction, while the wind model shows distinct epochs of higher and lower values. This is strongly correlated with the accretion rate --- higher rates result in less mass loss and vice versa, so since the accretion rate is above $\sim2\times10^{-8}\rmModot \rm{yr}^{-1}$ for most of the eruptions, hence the ratio ${m\rm_{ej}}/{m\rm_{acc}}$ is below the dotted line for most of the eruptions, meaning most of the eruptions in the wind model result in net mass growth.

\begin{figure}%[!ht]%
	\begin{center}
		{\includegraphics[trim={0.0cm 0.2cm 0.0cm 1.2cm},clip ,
			width=0.95\columnwidth]{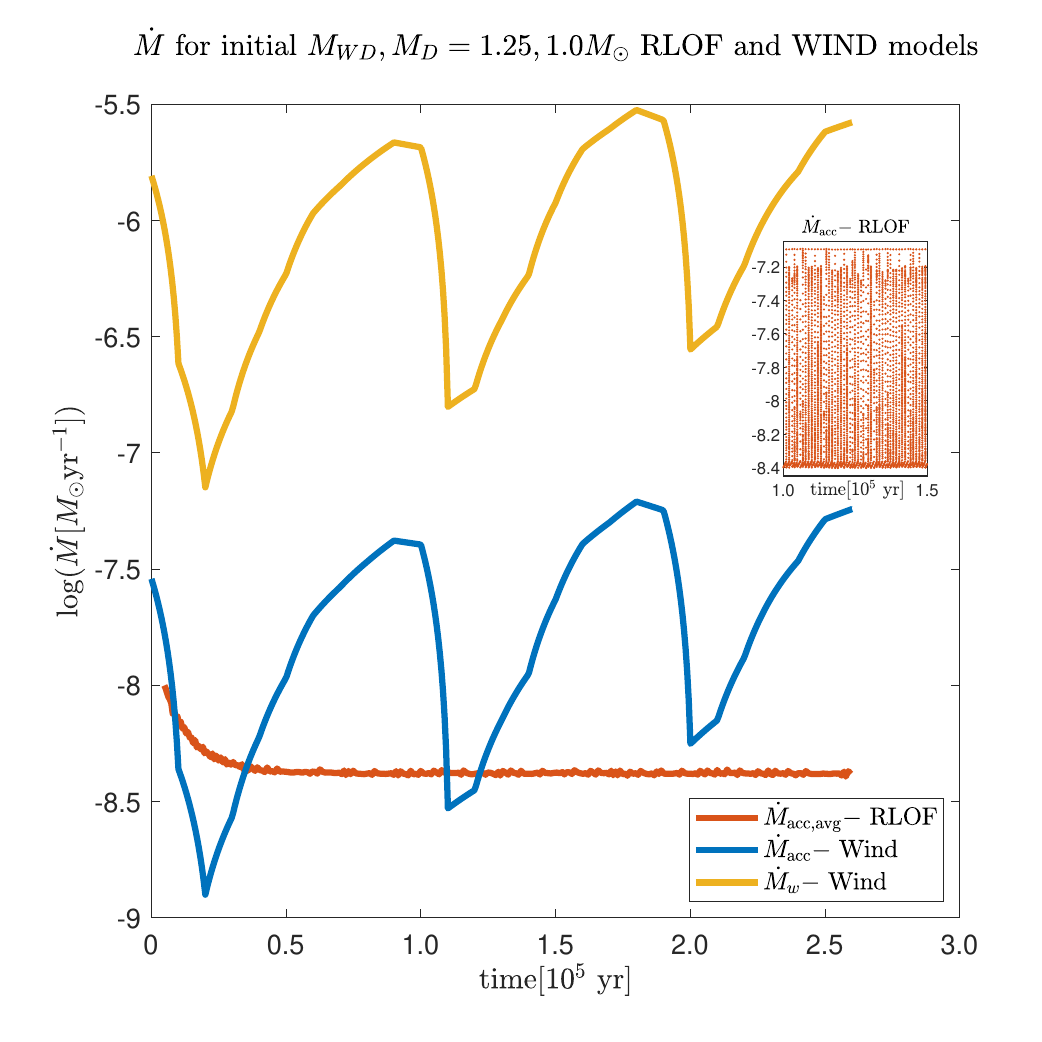}}
			\caption{The evolution of the wind rate (yellow); the accretion rate onto the WD in the wind model (blue); and the average accretion rate onto the WD in the RLOF model (red). The insert shows a portion of the RLOF model evolution, demonstrating the un-averaged behavior of the accretion rate in the RLOF model.	\label{fig:mdots}}
	\end{center}
\end{figure}

Figure \ref{fig:mdots} shows the very different manners in which the accretion rates ($\dot{M}\rm_{acc}$) of the two models evolve. For the RLOF model we show the average $\dot{M}\rm_{acc}$ per cycle, since it varies $\sim1.5$ orders of magnitude on a cyclic scale (shown in the insert in Figure \ref{fig:mdots}). The wind model does not experience this type of changes. The reason for this sort of variation in the $\dot{M}\rm_{acc}$ of the RLOF model is due to its nature, of being a close, Roche-lobe overflowing binary. The $\dot{M}\rm_{acc}$ starts out high due to irradiation from the previous nova eruption that occurred on its close WD companion. The irradiation causes temporary swelling of the donor envelope, increasing the RLOF and thus the mass transfer rate that then decreases over $\sim10^2$ years as the donor envelope cools and shrinks. It then very slowly increases again due to the decrease in the separation caused by angular momentum loss (AML) due to magnetic breaking (MB) and gravitational radiation (GR) \cite[]{Paxton2015,Hillman2020}. For simplicity, since the changing $\dot{M}\rm_{acc}$ over a single cycle yields similar results to taking a constant $\dot{M}\rm_{acc}$ that is equal to the average $\dot{M}\rm_{acc}$,  we show in Figure \ref{fig:mdots} the average $\dot{M}\rm_{acc}$ per cycle, which is fairly constant, at a rate of $\sim4\times10^{-9}\rmModot \rm{yr}^{-1}$ with the exception of the first few cycles which are still affected by initial conditions. 

The $\dot{M}\rm_{acc}$ onto the WD of the wind model experiences substantial changes, but not on a cyclic timescale. It is strongly linked to the wind rate from the AGB as seen clearly in Figure \ref{fig:mdots}. The wind rate takes this form due to it being correlated with the thermal pulses that characterize AGB stars, as explained in \S\ref{sec:intro}. The WD in this model experiences epochs of high accretion rate, and of low accretion rate, secularly varying, as dictated from Equation \ref{eq:mdotacc}. The reason that the accretion rate of the wind model does not feature large cyclic variation as seen in the RLOF model, even though the separation in this model decreases faster than in the RLOF model, is because when a system is in RLOF, every slight change in separation has a substantial effect on the mass transfer rate \cite[]{Ritter1988}. This becomes ineffective when the systems are detached. The range of accretion rates in the wind model include epochs with values higher and lower than $\sim2\times10^{-8}\rmModot \rm{yr}^{-1}$ which is (as explained above) the approximate limit for which above it a nova eruption will result in net mass growth and below it a nova eruption will result in net mass loss. This may be seen in Figure \ref{fig:masses} where the AGB star has epochs of decreasing and increasing mass, whereas the mass of the MS in the RLOF model only decreases. This is due to the average accretion rate of the RLOF model being fairly constant and a good deal below the limit throughout the simulation. It may also be seen in Figure \ref{fig:mej_over_macc} where the wind model has epochs below and above '1', whereas the RLOF model is almost always above '1'. 

\begin{figure}%[!hbt]%
	\begin{center}
	{\includegraphics[trim={0.1cm 0.5cm 0.0cm 1.4cm},clip ,			width=1.0\columnwidth]{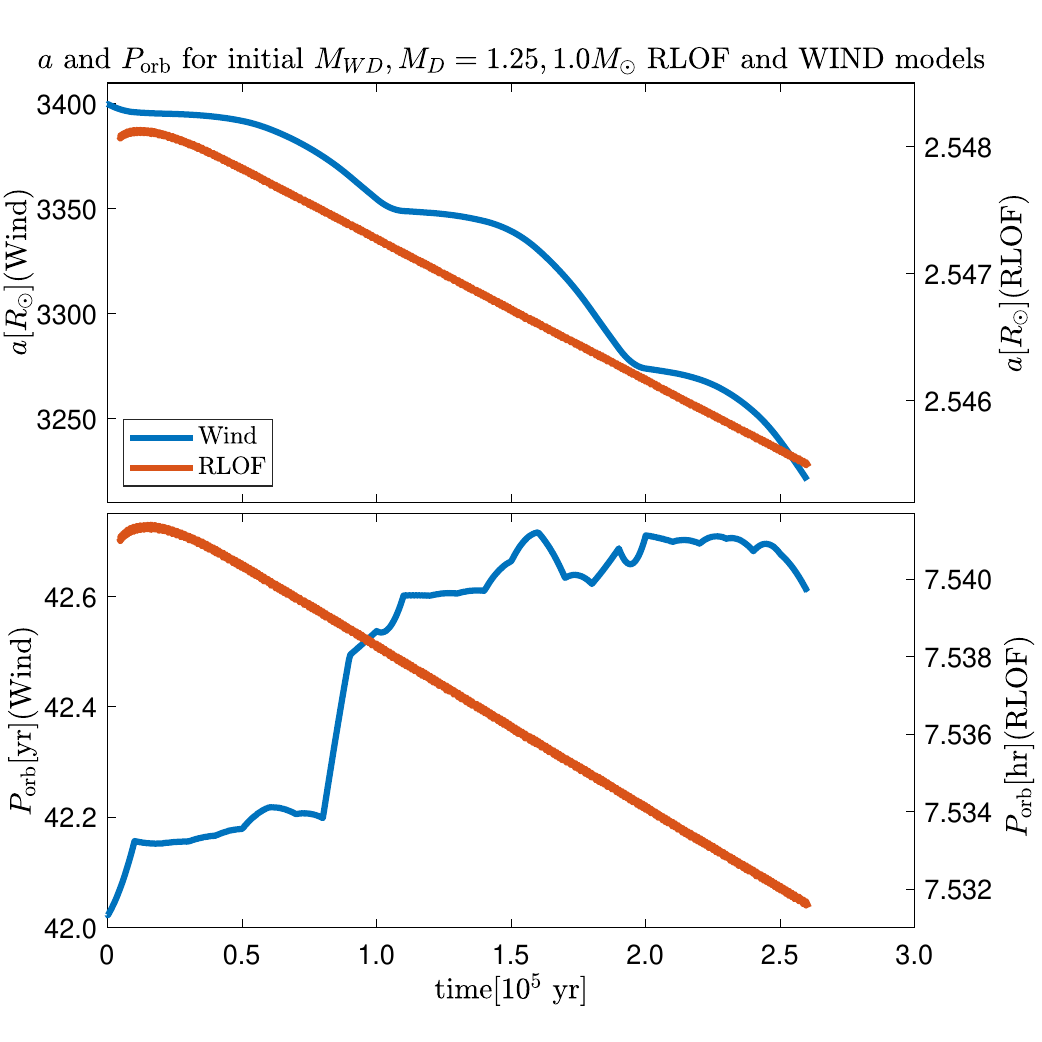}}
		\caption{The evolution of the separation (top) and the orbital period (bottom) for the wind model (blue) and the RLOF model (red). \label{fig:sep}}
	\end{center}
\end{figure}

In Figure \ref{fig:sep} we show the decrease in separation ($a$) and the consequential change in the orbital period ($P\rm_{orb}$). For the RLOF model $P\rm_{orb}$ decreases monotonically with the separation, dictated by Kepler's third law ${P{\rm_{orb}}=\left({4\pi G a^3}/({M_{WD}{+}M_D})\right)^{0.5}}$. For RLOF CVs the total binary mass ($M_{WD}{+}M_D$) changes much slower than $a^3$, therefore the period decreases as the stars move closer together. However, for systems dominated by mass loss via wind, the total binary mass will not necessarily change slower then $a^3$. If the mass decreases faster than $a^3$, then $P\rm_{orb}$ will increase as the stars move closer together. 
The drag force that is inflicted on the WD --- one of the main forces responsible for AML and thus for causing the separation to decrease --- is weaker for wider separations, while the rapid mass loss rate ($\dot{M}_w$) from the AGB star does not depend on its distance from the WD (although it does have substantial fluctuations). In the wind model we present here, the separation is relatively wide. This together with the fluctuating $\dot{M}_w$, yields epochs of $\Delta(M_{WD}+M_D)$ being larger than $\Delta(a^3)$ as well as vice versa, thus resulting in the $P\rm_{orb}$ decreasing and increasing in what only \textit{seems} like the sporadic behavior, exhibited in Figure \ref{fig:sep}.

\subsection{Eruption}\label{sec:eruption}
\begin{figure}%[!ht]%
	\begin{center}
		{\includegraphics[trim={0.4cm 0.5cm 1.2cm 1.1cm},clip ,
			width=0.95\columnwidth]{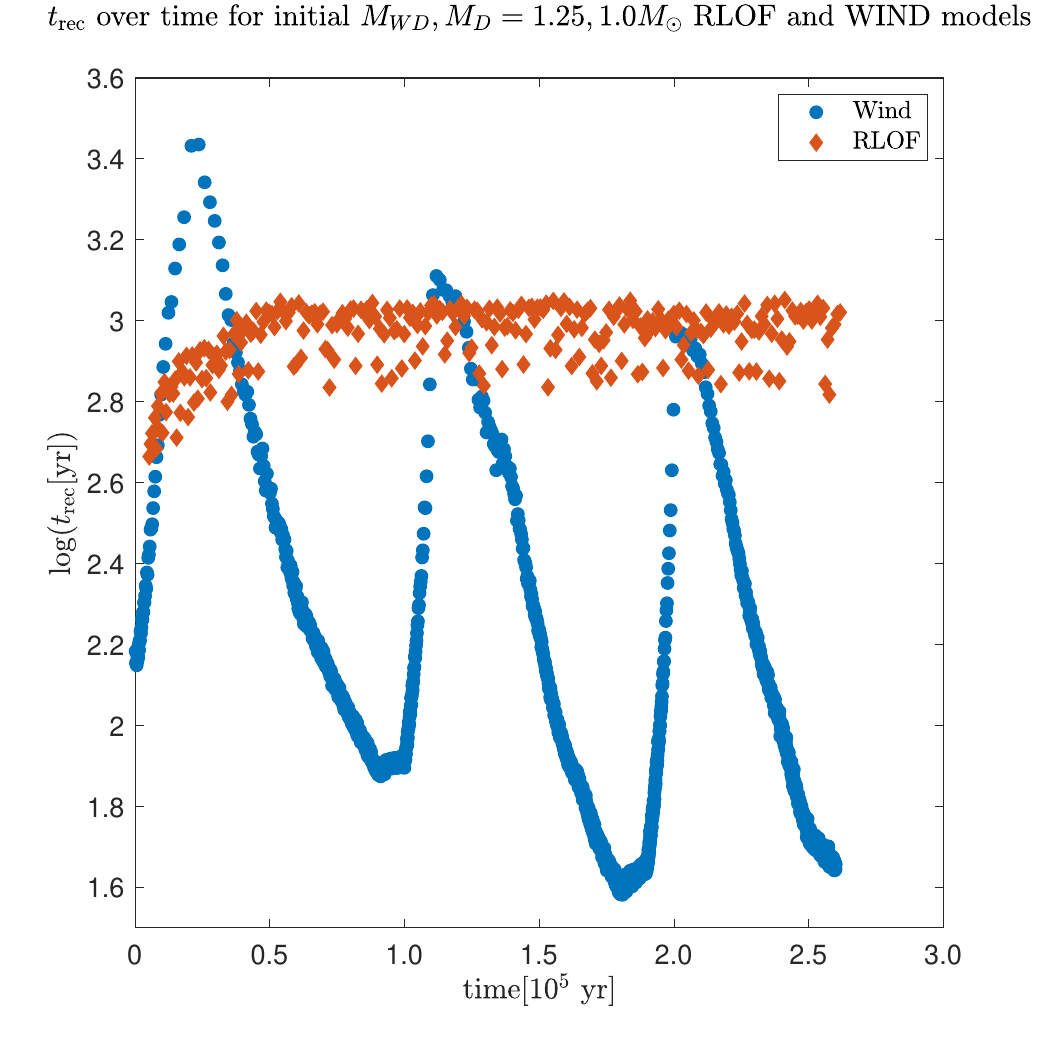}}
		\caption{The evolution of the recurrence period for the wind model (blue) and the RLOF model (red).		 \label{fig:trec}}
	\end{center}
\end{figure}

As explained above, during a period of $\sim2.6\times10^5$ years, the WD of the wind model produces {$\sim$7} times more nova eruptions then the WD of the RLOF model. This implies that the cycles (i.e., the accretion phase) of the RLOF model are $\sim$7 times longer than for the wind model. However, the evolution of the recurrence period ($t\rm_{rec}$), as shown in Figure \ref{fig:trec}, is more complex than this. The time between eruptions (i.e., the recurrence period, $t\rm_{rec}$) for the RLOF model is fairly constant, fluctuating within the range $\sim6\times10^2-1.2\times10^3$ years while for the wind model it may be as short as $\sim40$ years or as long as $\sim2.8\times10^3$ years --- a range of a factor of two as apposed to a factor of $\sim70$. 
 The behavior of $t\rm_{rec}$ is directly linked to the accretion rate, as clearly seen from the resemblance in the shape of the wind model curves --- when the average $\dot{M}\rm_{acc}$ (Figure \ref{fig:mdots}) over a single cycle of accretion is high, the $t\rm_{rec}$ is short and vice versa. For RLOF models, this correlation is supported by many authors, e.g., \cite{Prikov1995,Starrfield2012a,Wolf2013,Hillman2015,Hillman2016}. Here we find that this correlation applies for wind models as well. The matter ejected in the wind of an AGB star may be enriched in helium and heavy elements (mainly carbon) due to enhanced convection during the thermal pulses. Certain features of the nova (such as ejecta enrichment, the total amount of accreted mass ($m\rm_{acc}$) that is needed to trigger a TNR and the time that is required to accrete this mass ($t_{\rm rec}$), may be altered compared to a nova eruption as a result of the accretion of solar abundant matter. To assess the extent of this alteration we compared between two identical models ($M_{WD}=1.25M_\odot$ and $\dot{M}=10^{-9}M_\odot {\rm yr^{-1}}$) with the only difference being the abundance of the accreted matter --- one model consisting of heavy elements of solar abundance and the other of heavy elements enriched to double that of solar abundance. We found differences in various nova features of up to $\sim6\%$ between the two models. Noting that a more extensive examination may be informative (and we intend to explore this in a future study), we conclude that it is plausible that a wind-induced nova on the surface of a certain WD mass, accreting matter enriched in carbon at a certain average accretion rate, may not be \textit{identical} in behavior to a CV RLOF nova of the \textit{exact} same WD mass and average accretion rate, but rather it would resemble a CV RLOF nova eruption with a slightly different, of order a few percent, WD mass and/or accretion rate.
 
\begin{figure}%[!ht]%
	\begin{center}
		{\includegraphics[trim={0.4cm 0.5cm 1.2cm 1.1cm},clip ,
			width=0.95\columnwidth]{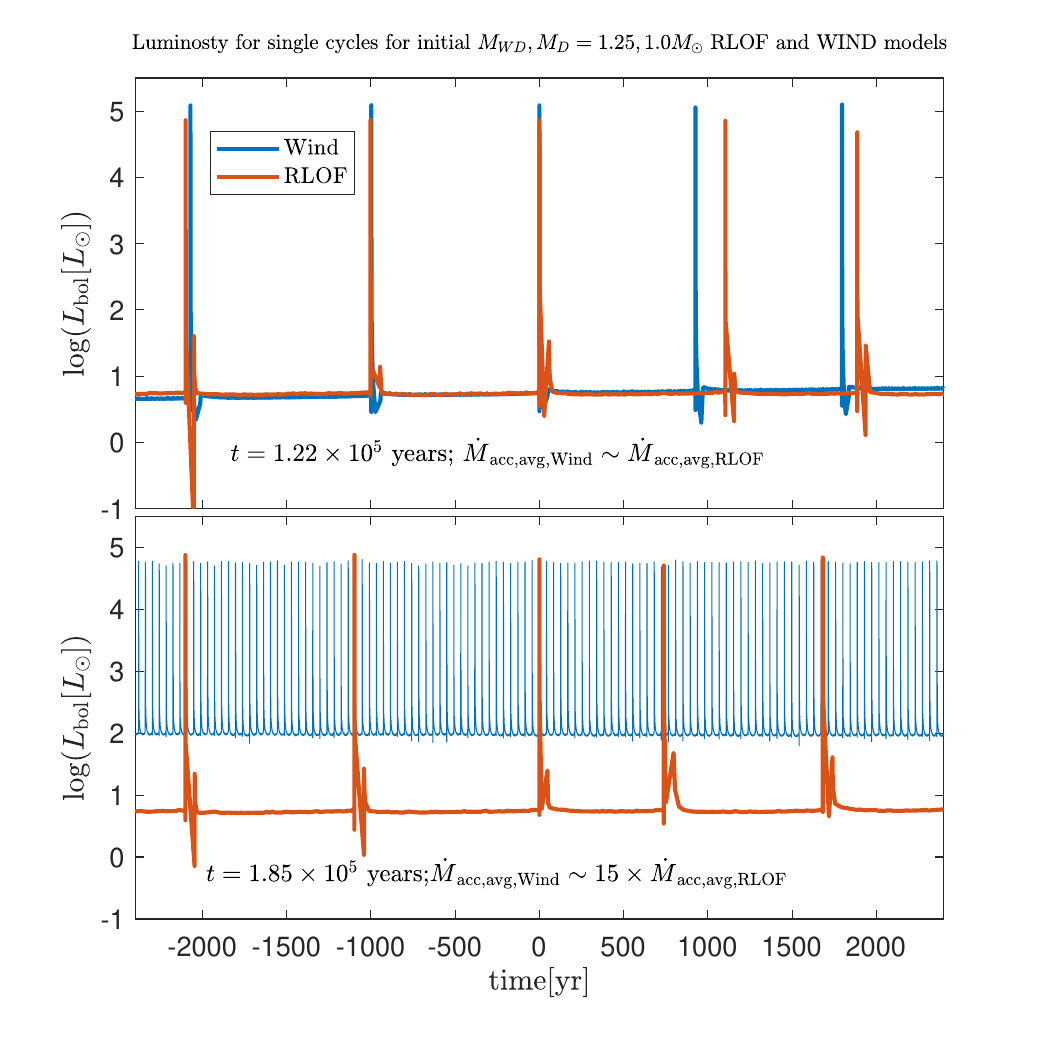}}
		\caption{Bolometric luminosity for a span of $\sim4000$ years at two different epochs. Top: at $t\simeq1.22\times10^5$ years for which the $\dot{M}{\rm_{acc,avg,Wind}}\simeq\dot{M}{\rm_{acc,avg,RLOF}}\simeq4\times10^{-9}\rmModot \rm{yr}^{-1}$; and bottom: at $t\simeq1.85\times10^5$ years for which the $\dot{M}{\rm_{acc,avg,RLOF}}$ is as above, while $\dot{M}{\rm_{acc,avg,Wind}}\simeq6\times10^{-8}\rmModot \rm{yr}^{-1}$. \label{fig:Lbol}}
	\end{center}
\end{figure}

Comparing between small groups of cycles from each model demonstrates the dominance of the accretion rate on the development of a nova, and that it is irrelevant whether the nova was induced by RLOF or by wind accretion. In Figure \ref{fig:Lbol} we show the comparison of two such sets of groups of cycles --- one at similar average accretion rates and the other at very different rates. The average accretion rate per cycle for the RLOF model is $\sim4\times10^{-9}\rmModot \rm{yr}^{-1}$ throughout the simulation, hence this is the average accretion rate for the RLOF model cycles sampled in both groups and the accretion rate for the wind model cycles in the top panel of Figure \ref{fig:Lbol}. The  accretion rate for the wind model cycles sampled in the bottom panel of Figure \ref{fig:Lbol} is about 15 times higher, $\sim6\times10^{-8}\rmModot \rm{yr}^{-1}$. The samples in each panel are taken at the same elapsed time from the beginning of the simulations spanning $\sim4000$ years, centered on a cycle chosen at about this time. The RLOF model cycles in the top and bottom panels and the wind model cycles in the top panel are indistinguishable from each other. They have the same recurrence time of $t\rm_{rec}\approx1000$ years, and amplitude of $A\approx10$ mag ($\approx4$ orders of magnitude difference between maximum and minimum luminosities), whereas the wind model cycles in the bottom panel experience $\sim25$ times more eruptions --- {$t\rm_{rec}\approx40$} years. In addition, the amplitude of the wind model eruptions in the bottom panel is substantially smaller, $A\approx7$ mag with  
a minimum luminosity of about $\sim20$ times higher than for the lower $\dot{M}\rm_{acc}$. The reason for this is the shorter cooling time between eruptions and the amplitudes are consistent with similar accretions rates from modeling \cite[e.g.,][]{Yaron2005,Hillman2014}. 

\begin{figure}%[!ht]%
	\begin{center}
		{\includegraphics[trim={1.0cm 0.7cm 1.2cm 1.1cm},clip ,
			width=0.95\columnwidth]{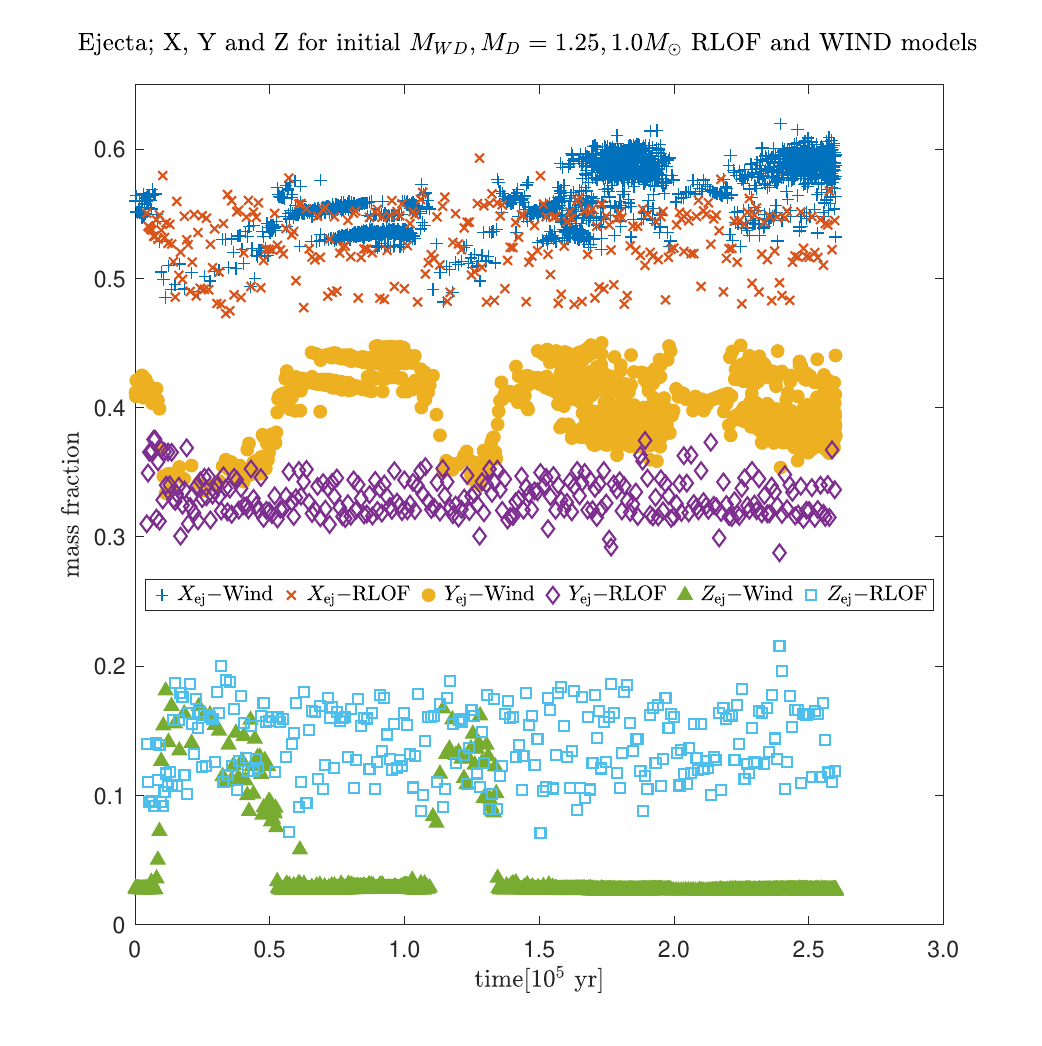}}
		\caption{Nova ejecta composition, showing mass fractions of hydrogen ($X$) helium ($Y$) and heavy elements ($Z$) for the wind model and the RLOF model. \label{fig:XYZ}}
	\end{center}
\end{figure}

The impact of the changing accretion rate is evident also from the composition of the mass ejected from the surface of the WD during a nova eruption, as may be seen in Figure \ref{fig:XYZ}. For the wind model cycles that have an average $\dot{M}\rm_{acc}$ similar to the average $\dot{M}\rm_{acc}$ of the RLOF model, the mass fractions of the two models overlap, for example, the heavy element abundance ($Z\rm_{ej}$) for these eruptions is roughly $0.10\lesssim~Z\rm_{ej}\lesssim0.18$. Then, as the $\dot{M}\rm_{acc}$ of the wind model increases, the mass fractions diverge. Following the heavy element mass fraction of the wind model in Figure \ref{fig:XYZ} demonstrates that the higher the $\dot{M}\rm_{acc}$ the lower the mass fraction of heavy elements in the ejecta. Using the same example as above, the heavy element abundance for the wind model cycles that experienced a high accretion rate have a substantially lower heavy element abundance of $Z\rm_{ej}\sim0.027-0.028$. The reason that a faster accretion rate results in less enrichment of heavy elements in the ejecta is because the critical accreted mass is reached faster, thus allowing less time for diffusive mixing and burning. 

The amplitude, the recurrence period of the wind model in the lower panel, and the ejecta abundances are all consistent with data of a recurrent RLOF nova from \citeauthor{Yaron2005} (2005, tables 2 and 3). Namely, comparing with their 1.25$\rmModot$ WD model 
with an accretion rate between $10^{-7}$ and $10^{-8}\rmModot\rm{yr}^{-1}$, the $\sim7$ mag amplitude of our wind model is between their corresponding $\sim6.5$ and $\sim9$ mag amplitudes; 
the $\sim40$ year recurrence period of our wind model is between their corresponding $\sim19$ and $\sim380$ year recurrence periods;  
and the $\sim0.027-0.028$ heavy element ejecta abundance of our wind model is between their corresponding $\sim0.027$ and $\sim0.10$ heavy element ejecta abundances. 

The epochs in our wind model with a high accretion rate of $\dot{M}{\rm_{acc}}\approx6\times10^{-8}\rmModot\rm{yr}^{-1}$ and a recurrence period of $\sim40$ years also resemble recurrent novae with the same WD mass (1.25$\rmModot$), average high accretion rate and recurrence period from novae models in \cite{Hillman2016}. As may be seen from their Fig. 4, the combination of the above mentioned values of $\dot{M}\rm_{acc}$ and $t\rm_{rec}$ fit in between the curves representing the WD masses of 1.2 and 1.32 $\rmModot$, which is consistent with our 1.25~$\rmModot$ WD wind model.   

These results emphasize that the manner in which the mass is transferred from the donor to the WD is not relevant to the results, but rather it is the rate this mass is transferred, that ultimately determines the evolutionary path.

\section{Discussion}\label{sec:discussion}
\subsection{The accretion rate}\label{sec:discussion-mdot}
When examining the accretion rate onto the surface of the WD in the wind model, it appears from Figure \ref{fig:mdots} that $\dot{M}\rm_{acc}$ is a constant fraction of approximately one hundredth of $\dot{M}_w$ throughout the simulation. This ratio is consistent  with other models \cite[e.g.,][]{Perets2013}. Nevertheless, our ratio $\dot{M}{\rm_{acc}}/\dot{M}_w$ is not constant, but rather slowly changes, increasing from a ratio of $\sim1.8\%$ at the onset of the simulation to  a ratio of $\sim2.2\%$ at the end of the simulation. The reason for the decrease is dictated from combining Equations \ref{eq:mdotacc} and \ref{eq:rho_w}, showing that the ratio $\dot{M}{\rm_{acc}}/\dot{M}_w$ increases with decreasing separation. For the wind model presented here this change is small, due to the wide separation. A more moderate separation would yield a more substantial change, which may be seen by combining Equations \ref{eq:rho_w} and \ref{eq:Drag} yielding an increase in the drag force with decreasing separation.
We note that the ratio also depends on the wind velocity which has a more complex effect and is beyond the scope of this study. 

\subsection{CNe and RNe in symbiotic systems}\label{sec:discussion-systems}

We turn to discuss how our results can be compared with observations of symbiotic novae.
Of the known eruptive symbiotic systems, only a few are widely separated systems that produce classical or recurrent novae. \cite{Akras2019} list over 250 Galactic symbiotic stars catalogued as S-type or D-type, most of which host a WD and may be a symbiotic nova, such as V1016 Cyg which is discussed below.
The New Online Database of Symbiotic Variables\footnote{\url{http://astronomy.science.upjs.sk/symbiotics/}} \cite[]{Merc2019b} lists the currently known symbiotic novae classified as S-type and D-type symbiotic and recurrent symbiotic novae, some of which are included in the following discussion.

While studying the symbiotic system AG~Peg, \cite{Kenyon1984} defined one of the possible formations of a symbiotic system to likely be a WD with an accretion disk being fed from a late-type giant. Their calculations show that provided the accretion rate is low enough ($\lesssim10^{-8}\rmModot\rm{yr}^{-1}$) a TNR may occur on the surface of the WD, producing a CN. They further explained that the observational characteristics of such a system would differ from those of CNe in CVs, due to the proximity of the giant donor and its dense wind. Considering the above, they deduced that the spectroscopic development of AG Peg since its optical maximum in 1850 is consistent with it being a CN eruption in a relatively close symbiotic system. The donor in AG Peg underfills its Roche-lobe by at least a factor of two \cite[]{Kenyon1987} and the system has been confirmed to be a binary system consisting of a WD and a late-type giant with an orbital period of $\sim2$ years \cite[e.g.,][]{Fekel2000}. The system has erupted again in 2015, this time the estimated accretion rate being $\sim3\times10^{-7}\rmModot\rm{yr}^{-1}$ and the eruption exhibiting an entirely different behavior \cite[]{Skopal2017,Merc2019a}. This is consistent with the general behavior of our wind model, where the wind rate may change by an order of magnitude, thus changing the accretion rate, yielding eruptions with different characteristics.

BF~Cyg, yet another well studied symbiotic system composed of a WD \cite[]{Kenyon1984} and a giant \cite[]{Kenyon1987} with an orbital period of $\sim2$ years \cite[e.g.,][]{Mikolajewska1987}. The WD mass is estimated to be $\sim0.6\rmModot$ and the donor is concluded to be a fairly normal M giant \cite[e.g.,][]{Mikolajewska1989,Leibowitz2006}. RLOF for this object is somewhat controversial where calculations show that it is not filling its Roche-lobe \cite[]{Mikolajewska1989,Fekel2001} while \cite{Yudin2005} deduced via modeling that the giant must be filling its Roche-lobe in order to explain elliptic attributes in the light curve. BF Cyg is known to have erupted twice; once in 1890 and again in 2006 \cite[]{Munari2006}, both reported as an increase of ${\sim}4$ mag in luminosity and a slow decline, the 1890 eruption taking a century to return to its pre-eruptive luminosity \cite[]{Leibowitz2006,Siviero2012}. 
 
As an example of a RN in a symbiotic system (SyRN), we shortly discuss V745 Sco \cite[]{Drake2016}, for which three nova eruptions have been recorded; in 1937, 1989 and 2014. Its orbital period is $\sim1.5$ years \cite[]{Schaefer2010}, its binary separation is of order 300$\rmRodot$ and the radius of the donor is $\sim120\rmRodot$ \cite[]{Orlando2017}. The RL radius being a factor of two more than the stellar radius would mean that the systems are well detached \cite[]{Murset1999} so accretion onto the WD would not be due to RLOF but rather from the wind of a red giant, although, it is not certain that processes such as tidal forces would not distort the donor causing ellipsoidal variability that would allow for RLOF \cite[]{Mikolajewska2008}.    

V1016 Cyg is one of the few \textit{widely} separated systems that are classified as a symbiotic novae due to the thermonuclear outburst that it underwent in 1965 causing an optical increase of $\sim5$--$7$ mag. \cite{Kenyon1987} presumed V1016~Cyg to be a detached binary, although the many estimates of the orbital period range from 6 to 80$\pm$25 years and the masses are undetermined as well, \cite[]{Parimucha2000,Brocksopp2002,Arkhipova2008}. 

The symbiotic nova HM Sge erupted in 1975 rising by $\sim 6$ mag (optical) interpreted as a nova eruption resulting from a TNR on the WD component of $\sim0.7\rmModot$ triggered by the accumulation of matter on its surface \cite[]{Richards1999,Sanad2020}. \cite{Nussbaumer1988} and \cite{Nussbaumer1990} pointed out that the abundance ratios of this eruption resemble those of novae and not those of symbiotic stars. \cite{Sanad2020} mentioned that the spectral behavior differs from the UV spectral behavior of a CN due the different environment --- a thick wind from a giant star.  By using colliding wind models of the interacting wind from the giant and the ejecta from the WD, and assuming the giants mass to be $\sim2\rmModot$, \cite{Richards1999} deduced a binary separation of $\sim$25 AU, implying an orbital period of order 100 years for reasonable stellar masses \cite[]{Schild2001}. This means that any mass captured by the WD would be from the giant's wind since the separation is far too wide for RLOF, as in our wind model, and the result is a nova, induced by TNR at the WD surface, as in our model. 
 
 Another similar system is the symbiotic system PU Vul 
 which was discovered in 1979 due to its slow increase of 6--7 mag to maximum, during which the acquired spectra was consistent with a nova outburst, classifying it as a very slow nova \cite[]{Kenyon1986,Kolotilov1995,Murset1999}.
The WD mass of this system is relatively small $\sim0.55\rmModot$ \cite[]{Kenyon1986}, the donor is a normal red giant \cite[]{Murset1999} and the orbital period is $\sim13.5$ years \cite[]{Kolotilov1995}. For a total system mass of $\sim2\rmModot$ Kepler's third law dictates a binary separation of $\sim1.5\times10^3\rmRodot$, which is about 5-10 times larger than a normal RG's radius. meaning that this system is detached and the donor must be transferring mass via wind alone, with no RLOF.

V407 Cyg was discovered in 1936, classified as a nova outburst from which it declined over the following three years \cite[]{Hoffmeister1949}. 
The system is a symbiotic star, comprising a ${\sim}1.2\rmModot$ WD accreting from a ${\sim}1.0\rmModot$ Mira donor, with an orbital period of {${\sim}43$ years} \cite[e.g.,][]{Munari1990,Pan2015}, removing any possibility for RLOF accretion. Their  estimated calculated accretion rate for a WD mass of $1.0$ and $1.4\rmModot$ is $1.0$ and $2.0\times10^{-8}\rmModot\rm{yr}^{-1}$ respectively, which is consistent with our model results. This system erupted again in 2010, brighter than before \cite[]{Aliu2012}, as a typical, TNR powered, classical nova outburst. In fact, V407 Cyg had erupted in 1998 as well, however the 1936 and 1998 eruptions were not as bright as the 2010 eruption, and only the 2010 has been definitely designated as a true classical nova \cite[]{Munari2011,Chomiuk2012b,Esipov2015}. Our wind model parameters are consistent with the parameters of the V407 Cyg system, and our key result, that CNe may occur in such systems, is supported by the system's 2010 CN eruption.

\section{Conclusions}\label{sec:conclusions}

We showed via a self-consistent simulation that includes the major physical forces that may affect the binary separation, that under the right circumstances, a WD accreting wind expelled matter from an AGB star, may experience classical nova eruptions.
We discussed a handful of symbiotic systems, with various masses, separation and giant type. Of these systems, HM Sge, Pu Vul and V407 Cyg are examples of systems that our model may represent --- confirmed wide binaries in which a TNR induced nova occurred as a result of wind accretion from a giant. The results we obtained are consistent with the observations of these systems, in particular with V407 Cyg which contains a rather massive WD ($\sim1.2\rmModot$) and a Mira variable. 

The rarity of such systems lies in the accretion rate. The separation of most symbiotic novae is relatively small, with $P\rm_{orb}$ of order a couple years, thus the accretion rate, even if may be just from wind, will be higher, triggering nova-like outbursts. However, if the separation is sufficiently wide --- such as in V407 Cyg ---  there is no physical reason to prevent the development of CNe.

We showed for the first time via self-consistent modeling, that CN may occur in wide symbiotic binary systems. 
We compared our findings with a model of a RLOF nova in a CV, and showed that the results are in-distinguishable from epochs of the wind model simulation with a similar accretion rate. In addition to these epochs, the wind model exhibited epochs of high accretion rate, that did not resemble the RLOF model that we analyzed here. However, we have shown that these epochs of a high accretion rate and a short recurrence period bare a striking resemblance to models of \textit{recurrent} novae with the same WD mass, average high accretion rate and recurrence period. 
The possibility of wind accretion inducing RNe was theoretically expressed in the extensive analysis of RNe by \cite{Schaefer2010} which shows similar eruption features of RNe in CVs and for wind accretion, such as the symbiotic system V745~Sco mentioned above.

In addition, our findings show that the long-term behavior over tens or hundreds of cycles can be in-differentiable from a long-term, multiple nova erupting CV. The results presented here show that, for a given WD mass, the development depends solely on the rate that mass is accreted onto the WD. This means that as long as the accretion rate is within the boundaries that characterize classical (or recurrent) novae in Roche-lobe overflowing CVs, the evolution of novae via wind accretion in wide binaries will advance in the exact same manner.   

We conclude that assuming the transferred mass is of solar composition, for given WD mass and average accretion rate, the development and behavior of a single nova eruption is no different than that of a classical or recurrent nova in a CV, thus, other properties of the system, including the nature of the donor are irrelevant.

\section*{Acknowledgements}
We thank Jaroslav Merc for enlightening discussions on symbiotic novae, and an anonymous referee for helpful comments.
We gratefully acknowledge support from the Authority for Research \& Development and the chairman of the Department of Physics in Ariel University.

\section*{Data Availability}
The data underlying this article will be shared on reasonable request to the corresponding author.

\bibliographystyle{aasjournal}
\bibliography{main}

\label{lastpage}
\end{document}